# IN-SITU HYDROGEN CHARGING OF ZIRCONIUM POWDER TO STUDY ISOTHERMAL PRECIPITATION OF HYDRIDES AND DETERMINATION OF Zr-HYDRIDE CRYSTAL STRUCTURE

T. Maimaitiyili[1], A. Steuwer[2], J. Blomquist[1], B. Matthew[3], Z. Olivier[4],
J. Andrieux[5], C. Bjerken[1], R. Fabienne[4]

[1]Malmo University, Sweden;
[2]Max IV Laboratory, Sweden;
[3]Manchester University, United Kingdom;
[4]IRSN DPAM/SEMCA/LEC, France;
[5]Europian Synchrotron Radiation Facility, France

## ABSTRACT

Zirconium alloys are widely used in the nuclear industry because of their high strength, good corrosion resistance and low neutron absorption cross-section. However, zirconium has strong affinity for hydrogen which leads to hydrogen concentration build-up over time. It is well known that the formation of hydrides will degrade the material and leads to, for example, delayed hydride cracking during high burn up. Even though zirconium hydrides have been studied for several decades, there still remain some controversies regarding the formation mechanisms, exact crystal structure, and stability of various hydride phases. This study uses high resolution synchrotron radiation as a probing tool to observe the precipitation and dissolution of hydrides in highly pure zirconium powder during in-situ hydrogen charging. The experiment enabled the direct observation of the hydride formation and phase transformations. It, also, provided high quality data for crystal structure determination.

## INTRODUCTION

Because of good mechanical properties at high temperatures and high pressure, excellent corrosion resistance in high temperature water, and a low thermal neutron absorption cross section, the zirconium (Zr) alloys have become some of the best structural material candidates in nuclear industries. As illustrated in Fig.1a, Zr-alloys are mostly used as fuel rod cladding to hold the nuclear fuel pellets inside the reactor core and are surrounded by neutron moderator and coolant. Depending on type of nuclear power plant, either light ($H_2O$) or heavy water ($D_2O$) are most commonly used as moderator. During reactor operation the environment inside the reactor core becomes very harsh (250-350 °C, 7-15 MPa) and Zr-alloys undergo corrosion according to chemical reaction:

$$Zr\ (s) + 2H_2O(l) \rightarrow ZrO_2(s) + 2H_2(g).$$

This chemical reaction is also true for $D_2O$, and in this case deuterium gas ($D_2$) will form instead of $H_2$.

Some parts of the hydrogen released through corrosion are absorbed into the cladding material and eventually leads to the precipitation of hydrides once the solubility limit is exceeded. Zirconium has relatively high solubility of hydrogen at high temperature but very low at room temperature [1]. Since the hydride phases have larger volume-per-unit than the solid solution phase (α-Zr), the formation of hydrides will introduce internal stresses and may lead to delayed hydride cracking [2, 3] (Fig.1a). Thus, the formation of hydrides is a potential issue during extended fuel burn-up [4] and when reactors are taken off-line and cooled to ambient temperature. To extend the life span of claddings and avoid catastrophic failure, it is necessary to identify the nature of various Zr-hydride phases and their exact structures.

According to literature, there exist at least three hydride phases at ambient temperature under atmospheric pressure (Fig.1b) depending on hydrogen concentration and quenching rate [5, 6]. Phases reported in the Zr-H system include stable face centered cubic (FCC) $ZrH_{1.65}$ known as δ-Zr hydride phase which is of CaF2-structure type (Fm-3m, a=4.7783 Å), the face centered tetragonal (FCT) $ZrH_2$ known as ε-Zr hydride phase with ThH2 structure (I4/mmm, a=4.9689, c=4.4479 Å), and the metastable FCT γ-hydride phase with structure type ZrH (P42/n, a=4.592 Å, 4.970 Å). All these reported phases are tabulated in Table 1.

Despite that Zr-hydrides have been studied for several decades, still some mechanisms are not fully understood, owing to the particular nature of hydrogen and its high diffusivity even at low temperature. According to literature, the stability of various hydride phases depends on alloying elements, internal stresses and the hydrogen isotopes that are considered [7, 8]. However, no full picture is given and ambiguities exist. Many of these reported structures are based on different separate experiments carried out on ex-situ hydrided samples using different techniques and facilities. Hence, there might some experimental errors or discrepancies involved between different measurements. To our knowledge there have not been any in-situ hydrogen charging studies performed on Zr.

In this work, we have performed in-situ hydrogen loading experiment at the beam line ID15-B at the European Synchrotron Radiation Facility (ESRF) in Grenoble, France. An onsite high pressure/high temperature capillary system was used to hydride high purity Zr powder to obtain all five reported phases in the Zr-H system through one single setup. Rietveld analysis was performed to determine the crystal structure of various phases.

## EXPERIMENTAL SETUP

### Sample For In-situ Measurements

The powder sample that was used in the experiment was prepared in strict order: First, a commercial grade, pure zirconium powder (99.2% purity) with maximum particle size about 45 microns was purchased from Goodfellow Ltd., Huntingdon, England. Then, the zirconium powder was filled into a glass container inside a glove box in argon environment to prevent any sort of contamination or oxidation. To dissolve any hydrides that may have formed earlier or during preparation, the sealed cells which contain Zr powder were first baked at ~700 °C for ~7.5 hours, then at ~1000 °C for 5 hours until an ultra-high vacuum level was achieved.

### Hydrogen Charging and Data Collection

The angular dispersive diffraction setup available at ID15B, with a high energy X-ray beam (87 keV) and large area detector (Trixell Pixium4700, Thales) positioned far from the specimen (~ 1.2 meter), was used for data collection. In our work, we used 0.2 or 20 second data acquisition (with corresponding real temporal resolution of 0.5 and 40 seconds) based on the phase transformations observed. The wavelength of the X-ray beam was calibrated with a standard $LaB_6$ specimen to λ=0.14232 Å. During the whole measurement, the wavelength was kept constant.

The mass spectrometer installed was used to detect any dehydration during the post-charging heat treatments. Since the inner diameter of the capillaries that were used for sample holding is ~2 mm, we selected 0.3x0.3 mm as beam size to ensure good diffraction signals from the specimens. During hydrogen loading we observed that the hydrogen diffusion was very fast and the heat produced during transformation was far beyond the limit of the capillaries, consequently we had to control the loading pressure between 0.05-1MPa depending on phase transformation rate.

To find out the stability and formation mechanisms of hydride phases, the measurement started at room temperature with pure powder (Fig.2b) and continuously data was collected at various times as function of temperature. For heat treatment during heating and cooling we used 10 °C/min. Once the temperature reached 300 °C, the system was kept fixed at 3.5 hours until no phase transformation was seen. Then, the system was heated up to 613 °C and then annealed back to room temperature with same rate.

## RESULTS AND DISCUSSION

After measurement the raw spectra were integrated using Matlab [9] followed by crystal structures refinement using structure analysis software packages Topas-Academic [10] and GSAS [11].

As the diffraction pattern is a convolution of instrument, sample and background contributions, we first used the Pawley method on the spectrum of a standard $LaB_6$ sample to determine the instrument function. We then performed full Rietveld analysis of each diffraction pattern collected at various time. A selection of Rietveld fits of diffraction patterns collected before and after hydrogen charging are shown in Figs 2b and 2a, respectively.

From Fig.2b, one can see that in the beginning of charging we only have pure α-Zr phase. After we introduced hydrogen gas at 300 °C we saw that α phase suddenly transformed into δ-hydride and β-Zr phases (Fig.2). Then, the β phase quickly disappeared and only pure δ-hydride was left. Moving on, the δ phase transformed into hydrogen rich FCT ε-hydride phases (Fig.2). This whole transition form α-Zr phase to ε-hydride phases completed in matter of seconds.

Once after the powder was fully transformed into ε-hydride, we did not observe any phase changes for several hours. To find out the reversibility of these transformations, we stopped the hydrogen flow and raised the system temperature to 613 °C under vacuum and we found that there are only α-Zr and δ-hydride in the system. However, when we lowered to 300 °C we did not observe any ε-hydrides. In the end of annealing when we returned to room temperature we detected α-Zr, δ-hydride and γ-hydride as shown in Fig.2 and Fig.3.

The lattice parameters of α-Zr phase after post heat treatment, obtained in this study ($a$ = 3.2420 (5) Å and $c$ = 5.1664(5) Å) are slightly larger than reported. During heat treatment till complete phase transformation, the α-Zr phase showed close to linear thermal expansion behavior respect to temperature. After cooling from 613°C to room temperature, we did not get exactly the same or even similar lattice parameters for α-Zr phase; instead we got an expanded structure as shown in Table 1. The slightly larger values of the α-Zr phase at room temperature prior to annealing might be caused by previous hydrogen charging, where the dissolved hydride could have caused the powder grain size to be smaller than originally, which could lead to slightly larger unit cells (on average) since the surface/bulk ratio is larger. As shown in Fig.2b, in the end of treatment there are about 9.64% of α-Zr phase, 81.94% of δ-hydride phase and 8.86% of γ-hydride phase in the system. Since hydrides are now acting as matrix they will lead to negative strain on the α-Zr phase. This eventually could add to the observed value of unit cell parameters and make the unit cell relatively larger than the reported.

The lattice parameters of δ- and γ-hydrides after heat treatment are in good agreement with data published in the literature. There is quite a significant difference between published β-Zr phase lattice parameters and what we found in our studies. Despite that our measurement temperature 560 °C is lower than earlier published measured lattice parameters; we still got quite expanded structure.

Based on the Zr-H phase diagram [6], the β-Zr and the δ-hydride phase only co-exist above 550 ºC for hydrogen concentration between 37.5-56.7 at.%H. Thus, it is not possible to observe β+δ at 300 ºC. However, the calculated lattice parameters of δ-hydride that formed together with β-Zr phase and just after the β to δ transformation are marginally larger than the lattice parameters obtained by Singh et. al. [12] for 500 ºC. Furthermore, the lattice parameter of β-Zr phase is 4% larger than the reported lattice parameter for 863 ºC [6]. All together indicates that the actual temperature of the Zr powder was significantly higher than what was measured by the thermo couple connected to the sample holder for a couple of fraction of a second during the early stages of the reaction. We believe that such high temperature rise occurs because of the very high reactivity of a powder (compared with a solid) due to the higher surface-to-bulk ratio. The heat produced during the exothermic reactions of H-solvation and hydride formation would, thus, raise the local temperature of the powder significantly for a brief time and it was not recorded by the thermo couple situated some distant away on the sample holder.

## CONCLUSIONS

In-situ hydrogen charging studies were performed on high purity Zr at ID15-B in ESRF in order to identify the formation mechanisms, stability and crystal structure of various Zr-hydrides. The following are the general conclusions of this study:
- The present study has demonstrated the potential use of high energy synchrotron diffraction on in-situ measurement of phase transformation and identifications of the Zr-H system.
- By using a high pressure/high temperature gas loading capillary system at the ESRF together with high energy dispersive setup, it is found to be possible to observe the kinetic behavior of phase transformations in real time even as it, in one case, was completed in less than 0.2 seconds in the case of $\alpha \rightarrow \beta$.
- For the first time, in-situ hydrogen loading experiment was performed on Zr at a 3$^{rd}$ generation synchrotron X-ray radiation facility, and all phases presented in the Zr-H phase diagram are obtained from a single setup..
- The determined lattice parameters of both α- and β-Zr phases are slightly larger than values reported in the literature. Negative strain from the hydride matrix on the α-particles on the one hand and local temperature increase due to exothermic reaction on the other were given as a possible explanation for either case.
- The hydrogen rich ε-phase reversibly transforms to first pure δ-phase and then α-Zr + δ-hydride as a function of H/Zr ratio as $H_2(g)$ was degassed from the system at 600 °C.

## ACKNOWLEDGEMENTS

The ESRF is gratefully acknowledged for the provision of beam time. We are thankful to the Swedish Research Foundation (VR 2008-3844) for financial support.

Table **1**: Some known and calculated crystallographic information about Zr and Zr-hydrides.

| Phase | Structure | Space group | a(Å) | b(Å) | c(Å) | Temperature [°C] | Reference |
|---|---|---|---|---|---|---|---|
| α(Zr) | HCP | P63/mmc | 3.2316 | 3.2316 | 5.1475 | 25 | [6] |
| | | | 3.24205 | 3.24205 | 5.16645 | 25 | Current studies |
| | | | 3.25721 | 3.25721 | 5.19881 | 25* | |
| β(Zr) | BCC | Im-3m | 3.6090 | 3.6090 | 3.6090 | 863 | [6] |
| | | | 3.75788 | 3.75788 | 3.75788 | 300 | Current studies |
| δ($ZrH_{1.66}$) | FCC | Fm-3m | 4.7783 | 4.7803 | 4.7803 | 20 | [6] |
| | | | 4.8051 | 4.8051 | 4.8051 | 500 | [12] |
| | | | 4.84159 | 4.84159 | 4.84159 | 300 | Current studies |
| | | | 4.77854 | 4.77854 | 4.77854 | 25* | |
| ε($ZrH_2$) | FCT | I4/mmm | 4.9689 | 4.4689 | 4.4497 | 20 | [6] |
| | BCT | | 3.51743 | 3.51743 | 4.52469 | 300 | Current studies |
| γ(ZrH) | FCT | P42/n | 4.592 | 4.592 | 4.970 | 17 | [6] |
| | | | 4.59966 | 4.59966 | 4.98654 | 25* | Current studies |

Note: * means it is after annealed.

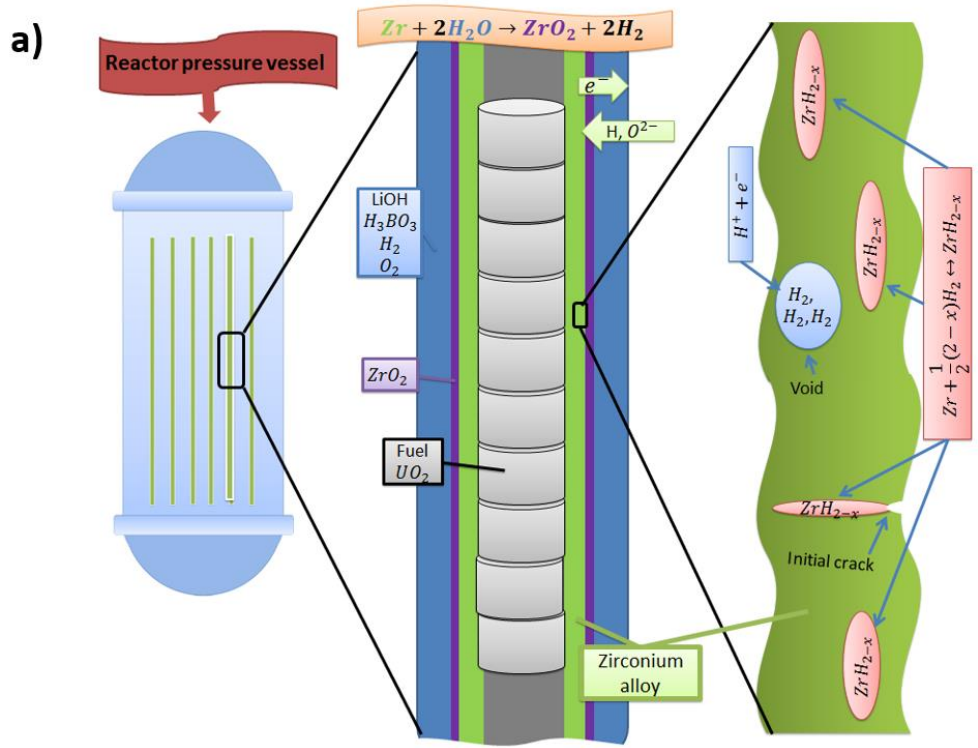

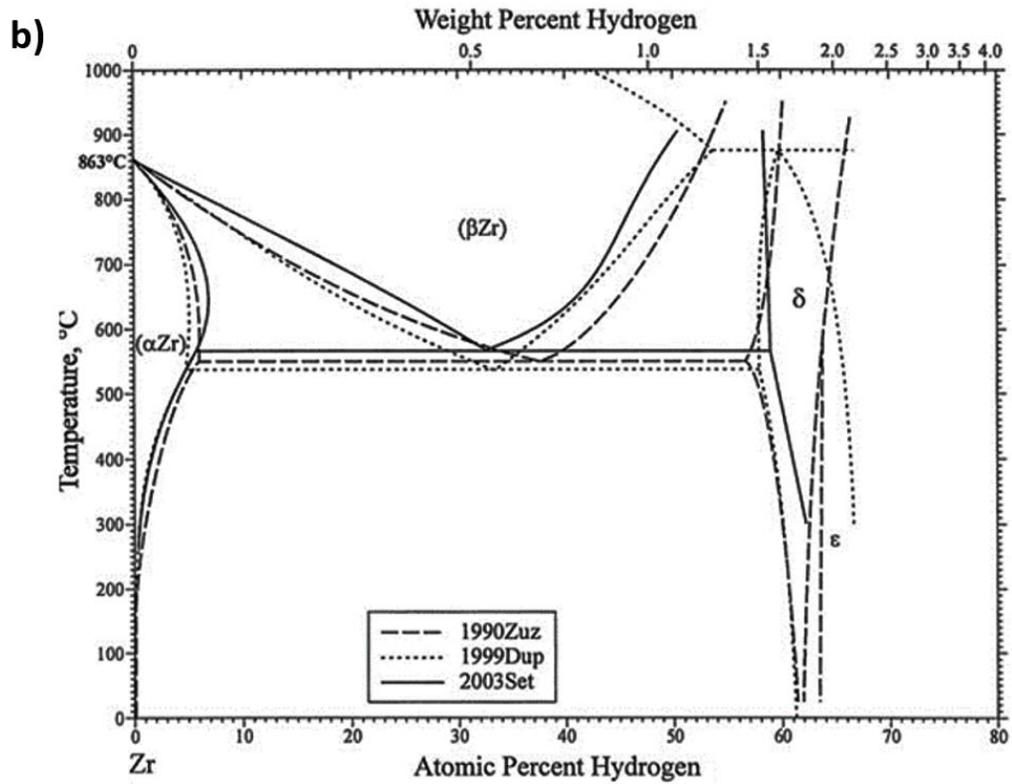

Figure 1: a) An illustration of a nuclear fuel cladding moderator system that may lead hydrogen related degradation; b) The Zr-H phase diagram [13].

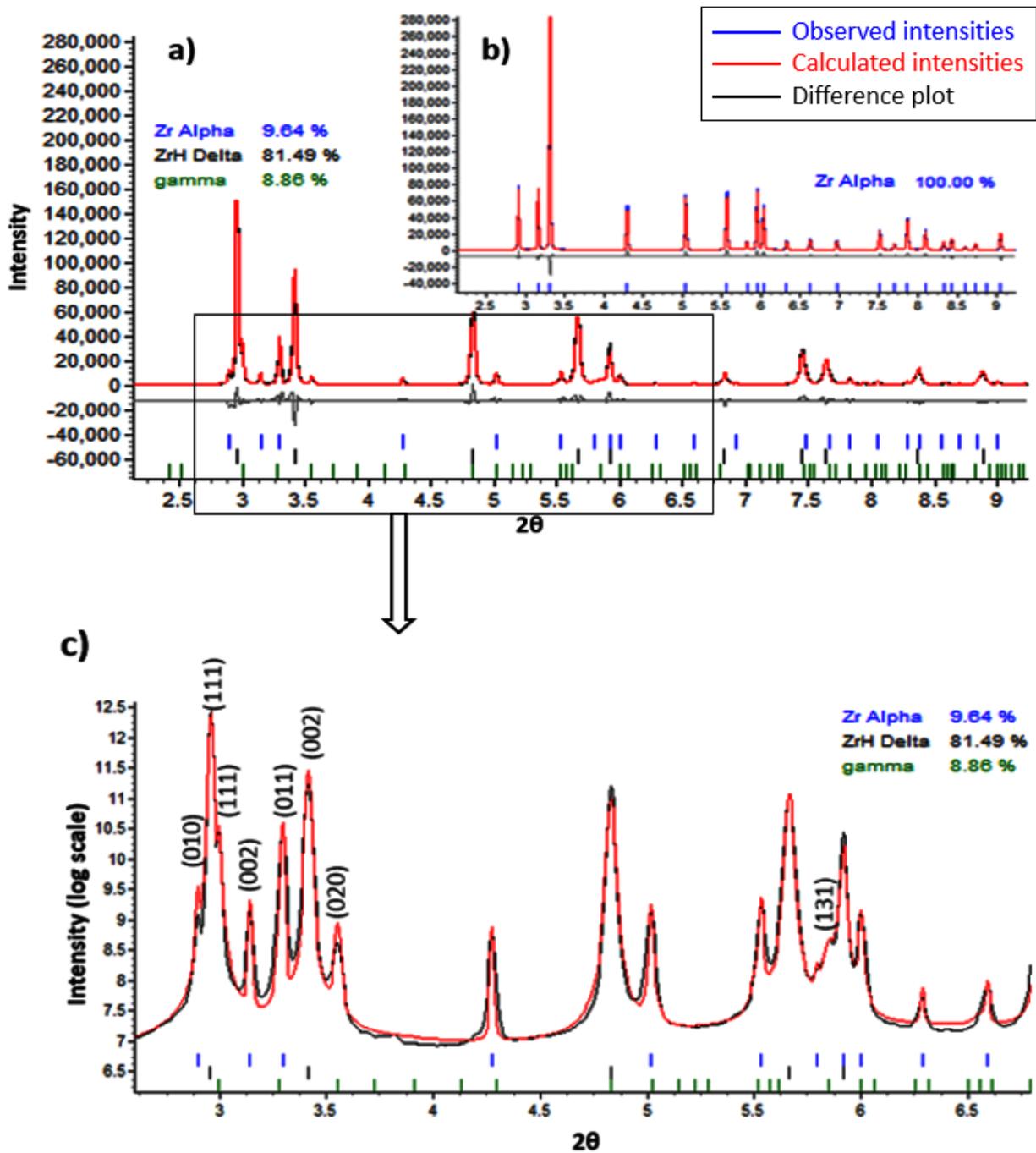

Figure 2: Typical diffraction patterns of the Zr powder before (b) and after (a) hydrogen charging which shows a typical Rietveld fit. The figure c) is the zoom of the indicated area in (a). Peak positions of various phases present are indicated by color coded tick marks under each peak.

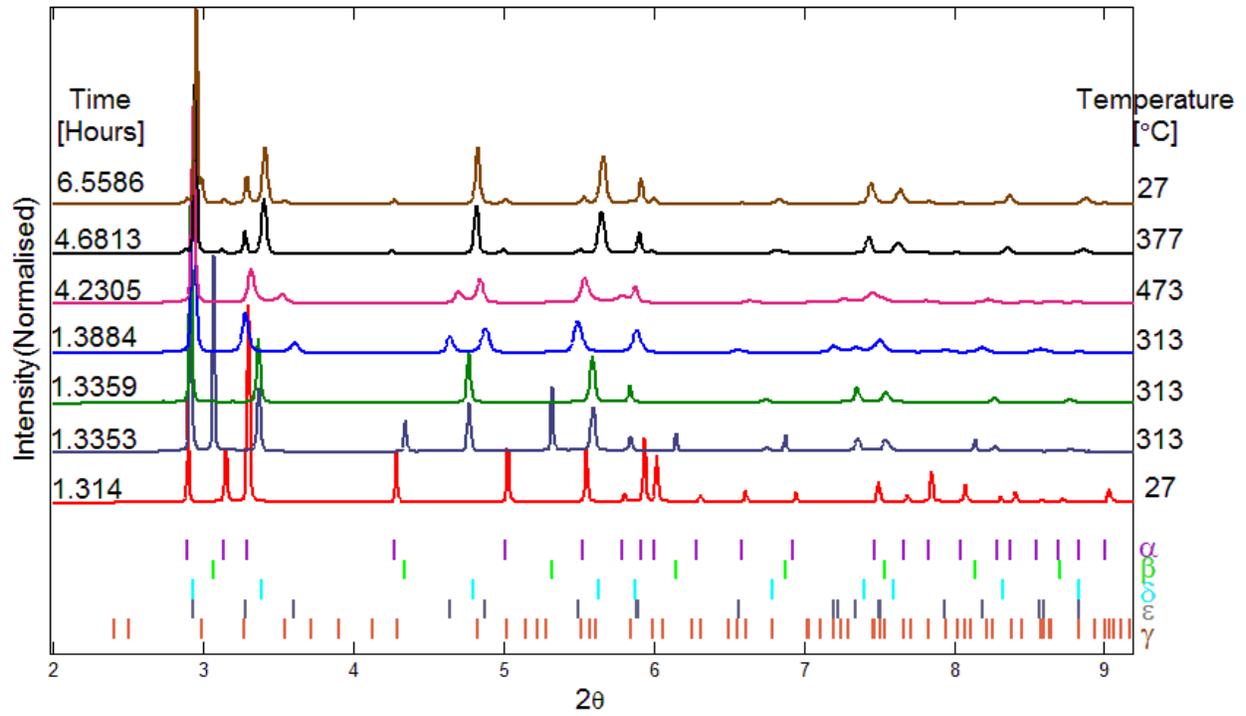

Figure 3: Selected diffraction patterns collected at different times and temperatures that shows various hydride and matrix phases.